\documentclass[amsfonts,amssymb,amsmath,twoside]{revtex4}
\begin{document}

\title{Statistical ensemble equivalence problem}
\thanks{Talk given at the conference \emph{''New Trends in High Energy Physics'', Crimea 2007}, Yalta, 15-22 September 2007}
\author{Ludwik Turko}
\email{turko@ift.uni.wroc.pl}
\affiliation{Institute of Theoretical Physics, University of Wroclaw,\\
Pl. Maksa Borna 9, 50-204  Wroc{\l}aw, Poland}
\date{November 3, 2007}

\begin{abstract}
A problem of the equivalence of statistical ensembles is critically analyzed. It is shown, that although different probability distributions of statistical physics have the same behavior in the thermodynamic limit, there are physical observables -- semi-intensive variables -- which keep memory of the underlying ensembles. This property is an universal one and can be observed even in the simplest case of the grand canonical and canonical ensembles of the classical statistical physics.
\end{abstract}
\maketitle
\section{Introduction}
There is a standard textbook knowledge about the equivalence of different statistical ensembles in the thermodynamic limit. It is usually understood that in the thermodynamic limit averages of intensive variables calculated with different ensembles have the same values. It's also seemed an obvious corollary that one gets the same limiting expressions of all relevant physical quantities. So it was quite surprising result \cite{begun 1} that scaled particle variance variables  defined as
\begin{equation}\label{cs_var}
    \omega = \frac{\langle N^2\rangle - \langle N\rangle^2}{\langle N\rangle}\,,
\end{equation}
with $N$ being the number of particles of a given charge, are different in the grand canonical, canonical  and microcanonical ensembles  -- even in the the thermodynamic limit. Corresponding ensembles considered there were canonical or grand canonical with respect to the conserved charge, while microcanonical means here an exact energy and charge value of the system. Those results were also repeated for a more realistic hadron-resonance gas \cite{begun 2}.

The statistical equivalence problem is not of academic interest only. Thermal statistical models are simple and effective tool to describe particle production in high energy heavy ion collision \cite{brs}. A given statistical ensemble is not a matter of an arbitrary choice but is related to internal properties of the system under consideration.  Different statistical ensembles for finite size systems lead to easily detected different predictions. However, in the case of ultrarelativistic heavy ion collision  departures from the thermodynamic limit can be ignored \cite{schr_pra}.

Results of \cite{begun 1, begun 2} triggered discussions about statistical equivalence problem. It was however rigorously shown \cite{crt1} that appropriate probability distributions have the same functional limit in the thermodynamic limit. This explain the notion of the statistical ensemble equivalence problem. It has also appeared \cite{crt2} that there is a general class of physical quantities which are finite in the thermodynamic limit but are still different for different statistical ensembles. Those quantities were named semi-intensive variables.  The scaled variance \eqref{cs_var} which was considered in \cite{begun 1,begun 2} belongs also to that class.

Let us consider here the simplest example of the standard statistical physics. It will be shown that even in this case the notion of semi--intensive quantities is relevant for the physical situation.

\section{Canonical and grand canonical ensembles of statistical physics}

The canonical ensemble is characterized by the fixed number of  particles $\bar{N}$, so the corresponding probability distribution is just Dirac delta function
\begin{equation}\label{C N distr}
    \mathbb{P}^C_{\bar{N}}(N,V) = \delta(N-\bar{N})\,.
\end{equation}
The grand canonical ensemble is characterized by the fixed averaged number of  particles $\langle N\rangle$. The number of particles is now a random variable. The  probability distribution to have $N$ particles in the system is denoted as $\mathcal{P}^{GC}_{\langle N\rangle}(N,V)$.

The thermodynamic limit is defined as the infinite volume limit taken under the condition that corresponding densities are kept constant. This means
\begin{equation}\label{C tlim}
    V\to\infty,\ \bar{N}\to\infty,\ \frac{\bar{N}}{V}=\bar{n}\,,
\end{equation}
for the canonical ensemble and
\begin{equation}\label{GC tlim}
    V\to\infty,\ \langle N\rangle\to\infty,\ \frac{\langle N\rangle}{V}=\langle n\rangle\,,
\end{equation}
for the grand canonical ensemble.

To formulate correctly  the thermodynamic limit of quantities involving densities, one should define  the  probability distributions for densities. They are
\begin{equation}\label{C prob dens}
    \mathcal{P}_{\bar{n}}^{C}(n,V)= V\mathbb{P}^C_{V\bar{n}}(Vn,V) = \delta(n-\bar{n})\,,
\end{equation}
for the canonical ensemble and
\begin{equation}\label{GC prob dens}
   \mathcal{P}^{GC}_{\langle n\rangle}(n,V)= V\mathbb{P}^{GC}_{V{\langle n\rangle}}(V n,V)\,,
\end{equation}
for the grand canonical ensemble.

The equivalence of the grand canonical and canonical distributions means
\begin{equation}\label{equiv}
  \mathcal{P}^{GC\infty}_{\langle n\rangle}(n)\equiv \lim_{V\to\infty}\mathcal{P}^{GC}_{\langle n\rangle}(n,V) = \mathcal{P}_{\bar{n}}^{C}(n)\,.
\end{equation}
In the thermodynamic limit the only relevant probabilities distributions are those related to densities. These distributions are expressed by probability moments. They are calculated for densities but not for particles. In the practice, however, we measure particles --- not densities as we do not know related volumes. Fortunately, volumes can be omitted by taking corresponding ratios.

Let us consider \emph{e.g.} the density variance $\Delta n^2$. This can be written as
\[\Delta n^2=\langle n^2\rangle - \langle n\rangle^2 =
\frac{\langle N^2\rangle - \langle N\rangle^2}{V^2}\,.\]%
By taking the relative variance
 \[\frac{\Delta n^2}{\langle n\rangle^2}=\frac{\langle N^2\rangle - \langle N\rangle^2}{\langle N\rangle^2}\,,\]
 volume-dependence vanishes.

A special care should be taken for calculations of ratios of particles moments. Although moments are extensive variables their ratios can be finite in the thermodynamic limit. These ratios are examples of semi-intensive variables. They are finite in the thermodynamic limit but those limits depend on volume terms in density probability distributions. One can say that semi-intensive variables ''keep memory'' where the thermodynamic limit is realized from.

Let consider as an example the scaled  particle variance
\[\frac{\langle N^2\rangle - \langle N\rangle^2}{\langle N\rangle}=
V\frac{\langle n^2\rangle - \langle n\rangle^2}{\langle n\rangle}\,.\]%

The term
\[\frac{\langle n^2\rangle - \langle n\rangle^2}{\langle n\rangle}\,.\]
tends to zero in the thermodynamic limit as $\mathcal{O}(V^{-1})$. So a behavior of
the scaled particle variance depends on the $\mathcal{O}(V^{-1})$ term in the scaled
density variance. A more detailed analysis of semi-intensive variables is given in
\cite{crt2}.

\section{Poisson distribution in the thermodynamic limit}
To clarify this approach let us consider a well known classical problem of Poisson distribution but taken in the thermodynamic limit.

Let us consider the grand canonical ensemble of noninteracting gas. A corresponding statistical operator is
  \begin{equation}\label{stat operator GC-P}
    \hat{D}=\frac{\,e^{-\beta\hat H+\gamma\hat N}}{\,\text{Tr}{\,e^{-\beta\hat H+\gamma\hat N}}}
\end{equation}

This leads to the partition function
\begin{equation}\label{GC-P part fn}
    \mathcal{Z}(V,T,\gamma)=\,e^{z\,e^\gamma}\,.
\end{equation}

where $z$ is  one-particle partition function
\begin{equation}
z(T,V)=\frac{V}{(2\pi)^3}\int d^3p\,\,e^{-\beta E(p)} \equiv V z_0(T)\,,
\end{equation}

A $\gamma$ parameter ($=\beta\mu$) is such to provide the given value of the average particle number $\langle N\rangle=V\langle n\rangle$. This means that
\begin{equation}\label{particle factor}
    \,e^{\gamma}= \frac{\langle n\rangle}{z_0}\,.
\end{equation}
The resulting probability distribution to obtain $N$ particles under condition that the average number of particles is $\langle N\rangle$ is equal to Poisson distribution
\begin{equation}\label{poiss}
    P_{\langle N\rangle}(N)=\frac{{\langle N\rangle}^N}{N!}\,e^{-\langle N\rangle}\,.
\end{equation}
We introduce corresponding probability distribution $\mathcal{P}$ for the particle number density $n=N/V$
\begin{equation}\label{probab dens}
\mathcal{P}_{\langle n\rangle}^{GC}(n;V)= V P_{V\langle n\rangle}(V n)=V\frac{(V\langle
n\rangle)^{V n}}{\Gamma(V n+1)} \,e^{-V\langle n\rangle}\,.
\end{equation}
The large volume behavior can be obtained here by means of the saddle point method. This gives
\begin{equation}\label{GC Vbehav}
    \mathcal{P}^{GC}_{\langle n\rangle}(n,V)\sim V^{1/2}\frac{1}{\sqrt{2\pi n}}
\left(\frac{\langle n\rangle}{ n}\right)^{V n} e^{V( n-\langle
n\rangle)}\left\{1-\frac{1}{12 V n}+\mathcal{O}(V^{-2})\right\}\,.
\end{equation}
This expression in singular in the $V\to\infty$ limit. To estimate a large volume behavior of the probability distribution \eqref{probab dens} one should take into account a generalized function limit. So we are going to calculate an expression

 \[\langle G\rangle_V=\int dn\, G(n)\mathcal{P}_{\langle n\rangle}^{GC}(n;V)\,,\]

for an arbitrary function $G(n)$ and with $\mathcal{P}_{\langle n\rangle}^{GC}(n;V)$ replaced by the asymptotic form from Eq \eqref{GC Vbehav}.

An asymptotic expansion of this integral is given by the classical Watson-Laplace theorem
\begin{quotation}
  \textbf{Theorem} Let $I=[a,b]$ be the finite interval such that
  \begin{enumerate}
    \item $\max\limits_{x\in I} S(x)$ is reached in the single point $x=x_0$, \mbox{$a<x_0<b$}.
    \item $f(x),S(x)\in C(I)$.
    \item $f(x), S(x)\in C^\infty$ in the vicinity of $x_0$, and $S^{''}(x_0)\neq 0$.
  \end{enumerate}
  Then, for $\lambda\to\infty,\ \lambda\in S_\epsilon$, there is an asymptotic expansion
\begin{subequations}\label{laplace}
  \begin{eqnarray}
    F[\lambda]&\thicksim &\,e^{\lambda S(x_0)}\sum\limits_{k=0}^\infty c_k\lambda^{-k-1/2}\,,
    \label{laplace main}\\
    c_k &=&\frac{\Gamma(k+1/2)}{(2k)!}\left(\frac{d}{dx}\right)^{2k}
\left.\left[f(x)\left(\frac{S(x_0)-S(x)}{(x-x_0)^2}\right)^{-k-1/2}\right]\right\vert_{x=x_0}\,.
\label{laplace coeff}
\end{eqnarray}
\end{subequations}
  $S_\epsilon$~is here a segment $|\arg z|\leqslant\frac{\pi}{2}-\epsilon<\frac{\pi}{2}$ in
the complex $z$-plane.
\end{quotation}

In the next to leading order asymptotic expansion one gets then
\begin{equation}\label{t lim 2}
    \langle G\rangle_V = G(\langle n\rangle) + \frac{\langle n\rangle}{2V}G^{''}(\langle n\rangle) +
    \mathcal{O}(V^{-2})\,,
\end{equation}
for any function $G$.

This gives us the exact expression for the density distribution \eqref{probab dens}in the large volume limit
\begin{equation}\label{poison t lim 2}
    \mathcal{P}_{\langle n\rangle}^{GC}(n;V) = \delta( n-\langle n\rangle)+\frac{\langle n\rangle}{2
    V}\,\delta^{''}( n-\langle n\rangle)+\mathcal{O}(V^{-2})\,.
\end{equation}
This confirms the equivalence of the grand canonical and canonical distributions as was defined in Eq \eqref{equiv}.

We are now able to obtain arbitrary density moments up to $\mathcal{O}(V^{-2})$ terms.
\begin{equation}\label{moments}
    \langle n^k \rangle_V = \int dn\, n^k \mathcal{P}_{\langle
    n\rangle}(n;V) = \langle n\rangle^k +\frac{k(k-1)}{2V}\langle
    n\rangle^{k-1}+\mathcal{O}(V^{-2})\,.
\end{equation}

We have for the second moment (intensive variable!)

\[\langle n^2 \rangle_V = \langle n\rangle^2 + \frac{\langle n\rangle}{V}+\mathcal{O}(V^{-2})\,.\]

This means
\begin{equation}\label{density limit}
    \Delta n^2=\frac{\langle n\rangle}{V}\to 0\,.
\end{equation}
as expected in the thermodynamic limit.

The particle number and its density are fixed in the canonical ensemble so corresponding variances are always equal to zero. The result \eqref{density limit} can be seen as an example of the equivalence of the canonical and grand canonical distribution in the thermodynamic limit. This equivalence is clearly visible from the Eq \eqref{poison t lim 2} where the delta function in the first term can be considered as the particle number density distribution in the canonical ensemble.

A more involved situation appears for particle number moments (extensive variable!). Eq \eqref{moments} translated to the particle number gives
\begin{equation}\label{particle moments 2}
        \langle N^k\rangle = \langle N\rangle^k + \frac{k(k-1)}{2}\langle
        N\rangle^{k-1}+\mathcal{O}(V^{k-2})\,,
\end{equation}

 One gets for the scaled variance (semi-intensive variable!)
\begin{equation}\label{scaled variance}
    \frac{\Delta N^2}{\langle N\rangle}=1\,,
\end{equation}
what should be compared with zero obtained for the canonical distribution.

The mechanism for such a seemingly unexpected behavior is quite obvious. The grand canonical and the canonical density probability distributions tend to the same thermodynamic limit. There are different however for any finite volume.
Semi-intensive variables depend on coefficients at those finite volume terms so they are different also in the thermodynamic limit.

Similar calculations can also be done for the energy taken in the canonical and in the grand canonical distribution \cite{trk1}. A conclusion is also similar. Energy probability distributions coincide in the thermodynamic limit for both ensembles, although finite volume corrections are different.

\subsection{Semi-intensive variables}
Now we are in position to create some semi-intensive variables. They are finite in the thermodynamic limit and have different values dependently on how the charge conservation is implemented in the description of the  system. There is  actually  a broad  class of variables. We take an an example
  \begin{equation}\label{particle mom sk}
    \mathcal{S}_k=\frac{\langle N^k\rangle - \langle N\rangle^k}{\langle N\rangle^{k-1}}\,.
\end{equation}
Taking into account Eq \eqref{particle moments 2} one gets in the thermodynamic limit
\[\mathcal{S}_k = \frac{k(k-1)}{2} + \mathcal{O}(V^{-1})\,.\]%
The scaled variance is just a  special case of $S_k$ corresponding to $k=2$.

Another examples are quantities closely related to cumulant or factorial cumulant moments or susceptibility ratios. Let define $p-$th order susceptibility \[\kappa_p=\frac{\partial^p\ln{\mathcal{Z}}}{\partial\mu^p}\,,\]%
where $\mathcal{Z}$ is a general partition function -- not only the Poissonian one.

One can easily check that the ratios
\begin{equation}
    \mathcal{K}_{p;r}=\frac{\kappa_p}{\kappa_r}\,,
\end{equation}
are semi--intensive quantities.

One can also construct more involved semi-inclusive variables having a finite thermodynamic limit behavior which are determined by higher order asymptotic terms of the corresponding probability distributions.

\section{Conclusions}
We have discussed  the differences in the asymptotic properties of the probability functions for a very simple classical system with both exact and average particle number conservation. We have shown that in the thermodynamic limit the corresponding probability distributions in the grand canonical and canonical ensembles coincide and are described as  generalized functions. This property is a direct consequence of the grand canonical and canonical ensemble equivalence in the thermodynamic limit. However, the first and higher finite volume corrections to the asymptotic value  differ in these ensembles.

Finally, we have derived the asymptotic behavior  of the particle moments using the results obtained for the probability functions. We have also applied these results to find the thermodynamic of  semi--intensive quantities. It was shown that in systems with exact and average particle number conservation such  quantities should naturally converge to different values in the thermodynamic limit. This is because the behavior of the semi--intensive quantities in the near vicinity to the thermodynamic limit are determined by the subleading, finite volume corrections to the probability distributions which are specific to a given statistical ensemble.

Semi--intensive variables are well suited to become benchmarks of different statistical distributions --- even in the thermodynamic limit. That property is an universal one and, as we've seen, it appears also on the simplest level. Is should be noted here that first moments of basic physical quantities tend to the same thermodynamic limit for different statistical ensembles. Equations of state are constructed as relations between first moments so they do not provide any information about underlying statistical ensembles. It can be achieved only by semi--intensive variables as \emph{e.g.} scaled variance or other fluctuations related quantities.
\begin{acknowledgments}
  It is my pleasure to dedicate this paper to Mark I.~Gorenstein on his 60th birthday.

This work was supported in part by the Polish Ministry of Science and  Higher Education under contract No. N N202 0953 33.
\end{acknowledgments}

\end{document}